\documentclass[pra,twoside,twocolumn,english,showpacs]{revtex4}
\usepackage{times}

\usepackage{graphicx}
\usepackage{amsmath}

\begin{document}

\title{Observing the Profile of an Atom Laser Beam}

\author{M. K\"ohl$^{1 \dag}$, Th. Busch$^{2,3}$, K.~M\o lmer$^3$, T. W.~H\"ansch$^{4,5}$, and T. Esslinger$^1$}
\affiliation{$^1$Institute of Quantum Electronics, ETH Z\"urich,
  CH--8093 Z\"urich, Switzerland\\
  $^2$Department of Physics, University College Cork, College Road,
  Cork, Ireland\\
  $^3$Department of Physics and Astronomy, University of Aarhus, DK--8000 Aarhus C, Denmark\\
  $^4$Sektion Physik, Ludwig-Maximilians-Universit\"at, Schellingstr.\ 4/III, D--80799 Munich, Germany\\
  $^5$Max-Planck-Institut f\"ur Quantenoptik, D--85748 Garching, Germany}

\date{\today{}}

\begin{abstract}
  We report on an investigation of the beam profile of an atom
  laser extracted from a magnetically trapped $^{87}$Rb
  Bose-Einstein condensate. The transverse momentum
  distribution is magnified by a curved mirror for matter waves and a
  momentum resolution of 1/60 of a photon recoil is obtained.
  We find the transverse momentum distribution to be determined
  by the mean-field potential of the residing condensate, which leads
  to a non-smooth transverse density distribution. Our
  experimental data are compared with a full 3D simulation of the output
  coupling process and we find good agreement.
\end{abstract}


\maketitle

Similar to the principle of an electron microscope, atomic matter
waves could be utilized to resolve structures on the nanometer
scale \cite{Balykin:87}. In contrast to electrons, atoms exhibit
an extremely small de-Broglie wavelength even at low energies.
This could allow for the gentle detection of nanoscopic textures
provided that atomic beams can be focused to a diffraction limited
spot size. Such an undertaking seems achievable with the
availability of Bose-Einstein condensates as nearly perfect matter
wave sources. From Bose-Einstein condensates well collimated and
bright atom laser beams have been extracted
\cite{Mewes1997,Anderson1998,Hagley1999,Bloch:99,Lecoq:01,Cennini:03,Robins:04}
which by far exceed thermal beams regarding spatial and temporal
coherence \cite{Andrews1997,Koehl:01,Ottl2005}.

Similar to laser operation in optics, the beam profile of an atom
laser is determined by the laser cavity and the beam extraction
from the cavity. Investigating the beam properties for coherent
matter waves is a difficult undertaking and so far no deviation
from the ideal behavior has been detected. Theoretically, the
longitudinal \cite{Schneider:99,Ballagh:00,Gerbier:01} and
transverse \cite{Busch:02} mode properties of beams have been
analyzed. Experimentally, the temporal coherence of atom lasers
\cite{Koehl:01} has been observed and the divergence angle
\cite{Lecoq:01} was measured, however without revealing the actual
beam profile. Here we present an experimental investigation of the
transverse momentum distribution of an atom laser beam and show
that it is influenced by the specific nature of the output
coupling mechanism.

A common method to extract atom laser beams from condensates in
magnetic traps is to continuously transfer a small fraction of the
trapped atoms into an untrapped state with a vanishing magnetic
moment \cite{Bloch:99,Lecoq:01,Robins:04}. The transferred atoms
no longer feel the confining trapping potential and are
accelerated by gravity, forming a downwards propagating beam.
Since the output coupling happens within the condensate, the beam
will initially experience the repulsive mean-field potential of
the residing condensate. This acts as an inhomogeneous refractive
index profile and will influence the transverse momentum
distribution \cite{Lecoq:01,Busch:02}.

\begin{figure}[htbp]
\includegraphics[width=.8\columnwidth]{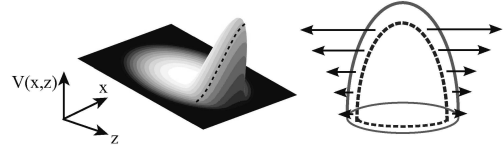}
\caption{Left: The combined inverted parabola potential caused  by
the trapped Bose-Einstein condensate and the gravity potential
acting on the $|F=1,m_F=0\rangle$ atoms in the atom laser beam.
The dashed line indicates the region where the resonance condition
is fulfilled and output coupling takes place. Right: Schematic
density profile of the trapped condensate (solid line) and the
output coupled atoms (dashed line). The arrows indicate the
transverse velocity that the output coupled atoms gain due to roll
off from the mean field potential of the remaining condensate.}
  \label{fig:Schematic}
\end{figure}

To quantify the influence of the residual condensate on the
transverse momentum distribution, let us consider a condensate in
the anisotropic harmonic trapping potential
$V_c=(m/2)[\omega_\perp^2(x^2+z^2)+\omega_y^2y^2]$ with the
z--axis oriented vertically. Since the force exerted by the
condensate on the beam is proportional to the gradient of the
condensate density, we will assume first that $\omega_\perp
>\omega_y$ and restrict ourselves to the two-dimensional situation
in the $x$-$z$ plane, where both trapping frequencies are high.
Upon transfer into an untrapped atomic state, the atoms are
subjected to the gravitational and the (repulsive) condensate
mean-field potential only. In the Thomas-Fermi approximation this
potential is given by
\begin{equation}
 \label{eq:Potential}
  V=mgz+\left\{
  \begin{array}{ll}
    \mu\left(1-\frac{x^2+z^2}{r_{\text{TF}}^2}\right)&
    {\text{if}}\; x^2+z^2<r_{\text{TF}}^2,\\
    0& {\text{else}},
  \end{array} \right.
\end{equation}
where $m$ denotes the atomic mass, $g$ the gravitational
acceleration, $\mu$ the chemical potential and $r_\text{TF}$ the
Thomas-Fermi radius of the strongly confined $x$- and
$z$-direction of the condensate. The output coupled atoms
experience a force proportional to the density gradient at their
respective position.

In the two dimensional setting the output coupling of the atoms
takes place on a circle of constant magnetic field, which is
defined by the resonance condition for the spin-flip transition.
Due to the gravitational force, however, the center of the
condensate is displaced from the minimum of the magnetic field by
an amount $z_0=g/\omega_\perp^2>r_\text{TF}$. In fact, the output
coupling region intersects with the condensate on an almost
horizontal slice. The forces due to the inhomogeneity of the
condensate density profile mainly affect the strongly confined
transverse direction (see Fig.~\ref{fig:Schematic}).

If the size of the repulsive mean-field potential would not be
finite, but simply an untruncated inverted harmonic oscillator
potential it would act as a diverging atom optical element
\cite{Lecoq:01}, since the atoms would always experience a force
proportional to their distance from the center of the potential.
We have recently argued that the truncation of the inverted
parabola potential of the condensate mean-field leads to
interference fringes in the transverse direction of the atom laser
beam \cite{Busch:02}. Two atoms starting at rest from {\em
different} transverse locations within the horizontal slice of the
output coupling region may end up with different velocities at the
{\em same} transverse position outside the condensate after a
certain time, which leads to quantum mechanical interference.
Therefore, the far-field distribution is not just simply a scaled
copy of the initial density distribution.

We have experimentally investigated the transverse momentum
distribution of an atom laser beam output coupled from a
magnetically trapped Bose-Einstein condensate. We start out by
producing Bose-Einstein condensates of about $5\times10^5$
$^{87}$Rb atoms in a Quadrupole-Ioffe configuration (QUIC) trap
\cite{Esslinger:98} in the $|F=1,m_F=-1\rangle$ hyperfine
groundstate by evaporative cooling.  The initial trapping
frequencies are $\omega_\perp=2\pi\times 110$\,Hz in the radial
and $\omega_y=2 \pi \times 14$\,Hz in the axial direction, so that
our condensate shape is cigar-shaped. After formation of the
condensate we increase the radial trapping frequency to
$\omega_\perp=2\pi\times 200$\,Hz over a time scale of 1 second.
Due to the magnetic field geometry of the QUIC-trap and the
nonlinear permeability of the $\mu$-metal enclosure the axial
confinement is weakened in this process leading to a trap
frequency of $\omega_y=2 \pi \times 11$\,Hz. The final chemical
potential is 2.5\,kHz and the Thomas-Fermi radius in the radial
direction is $3.7\,\mu$m. The atom laser beam is then extracted
from the condensate by continuous output coupling \cite{Bloch:99}:
a weak monochromatic radio-frequency field transfers atoms locally
from the magnetically trapped condensate into the untrapped state
$|F=1,m_F=0\rangle$ in which the atoms propagate downwards due to
gravity and form a collimated beam.

\begin{figure}[tbp]
\includegraphics[width=.8\columnwidth,clip=true]{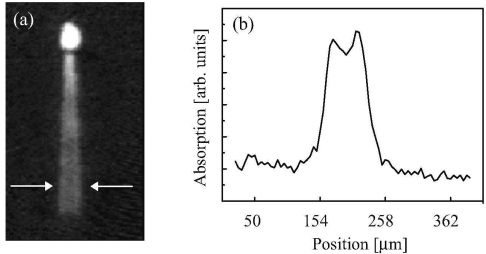}
\caption{{\bf(a)} Atom laser beam extracted from a compressed
Bose-Einstein condensate. The double peak structure of the beam is
clearly visible. The height of the image is 780\,$\mu$m. {\bf(b)}
Cut through the atomic density distribution at the position
indicated by the arrows.} \label{fig2}
\end{figure}

An absorption image of an atom laser beam is shown in
Fig.~\ref{fig2}(a). The line of sight is chosen along the symmetry
axis of the cigar shaped condensate. The atom laser beam exhibits
a two peak structure in the transverse direction. A horizontal cut
through the density distribution of the atom laser beam
500\,$\mu$m below the Bose-Einstein condensate clearly reveals
this structure (see Fig.~\ref{fig2}(b)). This indicates that the
transverse momentum distribution of the atom laser is not a scaled
copy of the condensate wave function and it is not sufficiently
well described by just a divergence angle. The difference compared
to the results reported in \cite{Lecoq:01} we attribute to the
experimental settings. The direction of observation in our
experiment is chosen along the symmetry axis of the condensate
whereas in \cite{Lecoq:01} the line of sight was at an angle of
$55^\circ$ with respect to the symmetry axis which might have
averaged out details of the density distribution. In the direction
of view orthogonal to the symmetry axis of the condensate we find
no modification of the atom laser density. This is consistent with
the substantially smaller curvature of the mean-field potential
along this axis. Moreover, in our experimental configuration the
confining magnetic potential is stronger and correspondingly the
mean-field curvature is more pronounced.

\begin{figure}[tbp]
\includegraphics[width=.8\columnwidth,clip=true]{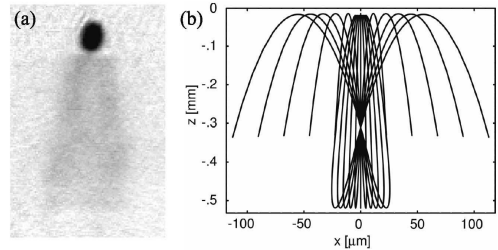}
\caption{{\bf (a)} Atom laser after reflection by the curved
mirror. The height of the image is 780\,$\mu$m and it is taken
33\,ms after starting the output coupling process including 3\,ms
ballistic expansion for the Bose-Einstein condensate. {\bf (b)}
Classical trajectories of atoms being reflected from the curved
matter wave mirror. After reflection the atoms laser is beam is
focused about $300\,\mu$m below the Bose-Einstein condensate. The
evolution is shown up to 30\,ms after the start of output
coupling.} \label{fig3}
\end{figure}

For a more detailed investigation of the structure of the beam a
higher momentum resolution in the transverse direction is
desirable. We accomplish this by employing a matter wave mirror in
a curved mirror configuration such that the mirror acts as a
``magnifying glass`` for the atom laser beam. The principle of the
atom mirror is to switch the interaction between the atoms in the
atom laser beam and an external magnetic field on and off by an
optically induced spin-flip transition \cite{Bloch:01}. A pair of
phase-coherent laser beams drive a two--photon hyperfine Raman
transition between the magnetically untrapped state
$|F=1,m_F=0\rangle$ and the magnetically trapped state
$|F=2,m_F=1\rangle$ approximately 500\,$\mu$m below the position
of the Bose-Einstein condensate. After the atoms are transferred
into the $|F=2,m_F=1\rangle$ state by adiabatic passage on their
way downwards they are reflected off the inhomogeneous field of
the magnetic trap. On their way upwards they pass through the
Raman lasers again and are coherently transferred back into the
initial state $|F=1,m_F=0\rangle$ after which they move
ballistically in the gravitational potential. The efficiency of
this reflection process is almost 100\% \cite{Bloch:01}, the
mirror preserves the coherence of the incident atom laser beam
\cite{Koehl:01} and it operates continuously, as opposed to curved
mirrors for Bose-Einstein condensates based on pulsed magnetic
fields \cite{Arnold2002}.

\begin{figure}[tbp]
\includegraphics[width=.85\columnwidth,clip=true]{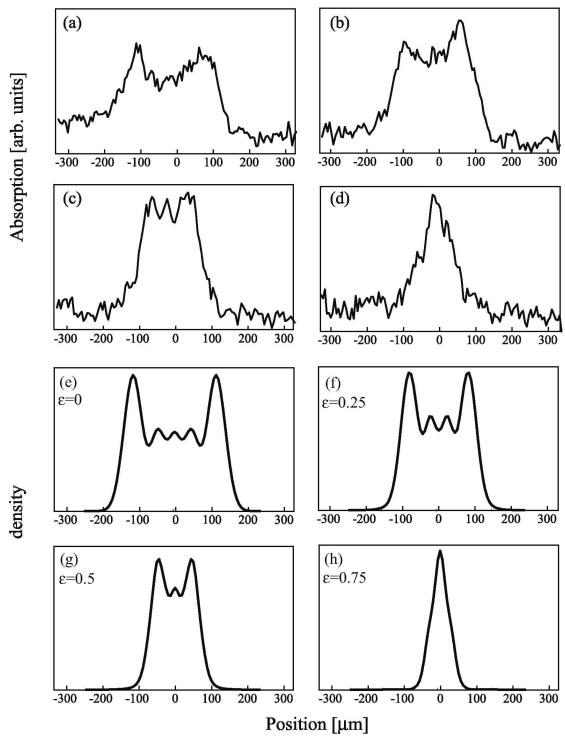}
\caption{(a)-(d): Transverse density distribution of atom laser
beam after momentum magnification by the curved mirror. The
profiles correspond to different output coupling regions within
the Bose-Einstein condensate (top (a) to bottom (d)). The
difference in the output coupling frequency is 2\,kHz between each
slice [(a)-(d)] corresponding to $1\,\mu$m-steps of the vertical
extraction point. The density distribution is evaluated
250\,$\mu$m below the position of the condensate. (e)-(h): Density
profiles from a full 3D calculation for various output coupling
positions $x_c=x_0-\epsilon r_\text{TF}$, corresponding
approximately to the experimental data. The density is averaged
along the long axis of the condensate, which takes into account
the effect of the absorption imaging. Displayed is the far field
density distribution as given by the Fourier transform of the
density well below the condensate. The momentum distribution has
been converted into  position space according to the classical
trajectories shown in figure \ref{fig3}b.} \label{fig4}
\end{figure}

The ''surface'' of the mirror is defined by the lines of constant
magnetic field, which are approximately concentric cylinders with
the condensate being displaced only slightly from their center. In
light optics one would expect a wave emerging from the center of
the curved mirror being reflected back onto itself. In matter wave
optics the situation is different: the gravitational acceleration
collimates the atom laser beam since after half a millimeter
dropping distance in the gravitational potential the longitudinal
velocity ($\approx$10\,cm/s) has already overcome the transverse
velocity spread by more than one order of magnitude. The reflected
atom laser beam is therefore focused at a distance about half the
radius of curvature away from the mirror (see Fig.~\ref{fig3}(b)).
Due to the axial symmetry of the magnetic trapping potential, the
focussing occurs only in one transverse direction. During the
reflection from the curved potential the transverse momentum
changes by a factor --2.6, as calculated from a simulation of the
classical trajectories. The atomic density distribution is imaged
by absorption imaging 20--30\,ms after the reflection of the atoms
and the beam is widened approximately by a factor of 6. The
amplification factor is determined by the longitudinal velocity
and the transverse location of the reflecting point of the atoms
which is proportional to the initial transverse velocity. It is
crucial for the application of this momentum magnification scheme
that the beam is already Fourier separated in the transverse
direction (i.e.~the transverse velocity is a unique function of
the transverse coordinate) when it enters the mirror. Otherwise
the curved mirror will modify the far field evolution of the atom
laser beam.

The achieved momentum resolution in the transverse direction is
very high: given the spatial resolution of the imaging system of
$5.2\,\mu$m, the momentum magnification by a factor 2.6 and
observation 20\,ms after the atoms were reflected off the mirror,
one obtains 0.1\,mm/s equivalent to 1/60 photon recoil.

The transverse atomic density distribution 20\,ms after reflection
from the Raman mirror is shown in Fig.~\ref{fig4}(a)-(d) for
different extraction regions inside the Bose-Einstein condensate.
The traces differ by 2\,kHz in output coupling frequency or
equivalently by $1\,\mu$m in output coupling position. For output
coupling from the bottom of the condensate (Fig.~\ref{fig4}(d))
the transverse momentum distribution is a single peaked function.
This results from the short interaction time of the atom laser
beam with the mean-field potential. When the extraction of the
atom laser beam is performed closer to the center of the
condensate the interaction time and the mean-field potential
increase which leads to a larger transverse momentum spread of the
beam. The apparent noise on the density profiles which masks finer
details of the interference fringes comes from the very low
densities of the expanded atom laser beam. The optical density is
lowered by the same factor of 6 as compared to conventional atom
laser beams as the momentum resolution is enhanced. Working at a
higher atomic flux or integrating over several repetitions of the
same experiment could improve on the signal to noise ratio.

We have compared the measured traces with a full, three
dimensional numerical simulation of the output coupling process.
Since the curvature of the mean--field--potential is comparatively
small along the axial direction, no significant dynamics in the
$y$-direction is observed as expected. At the same time the
dynamics in the $x$-direction strongly scales with the exact
location along the symmetry axis, due to the inhomogeneous density
of the condensate. While the distributions for a single, fixed
$y$-value shows an interference pattern with high visibility
\cite{Busch:02}, the integration along the line-of-sight averages
out most of these pronounced two--dimensional interference
fringes. In Fig.~\ref{fig4}(e)-(h) we show the calculated pattern
which show a good qualitative agreement with the experimentally
observed density distributions.

In the present configuration  after reflecting off the curved
mirror we obtain a very tight focus for the atom laser beam with a
waist below the resolution limit of detection optics. We focus the
atom laser beam with an f-number \cite{remark3} of approximately
3. With this low f-number and the inherently short de-Broglie
wavelength of a matter wave beam we estimate a focal spot size in
the 100\,nm regime and correspondingly a high atomic densitiy in
the focus may be obtained. Given the the repulsive interactions
between the rubidium atoms, a self-defocusing of the atom laser
beam due to mean-field repulsion could be expected. However, with
the present atomic density in the beam of about
$10^{11}$\,cm$^{-3}$ and for an (assumed) diffraction limited spot
size the density gradient is too small to lead to a significant
self-defocusing. We estimate that for an initial beam density of
$10^{14}$\,cm$^{-3}$ and focusing in both transverse directions
the self-defocusing effect would become comparable to the usual
mean-field expansion of a condensate, which could be easily
detected. This regime of nonlinear matter wave propagation at high
intensities would be very interesting to study since principal
analogies with the propagation of high-intensity laser beams in
nonlinear optical media exist. For atomic beams however, the
nonlinearity is an intrinsic property of the beam.

In conclusion, we have studied the influence of the mean field
potential of a Bose-Einstein condensate on the transverse momentum
distribution of an atom laser beam. Using a curved matter wave
mirror, we have magnified the transverse momentum distribution and
obtained a momentum resolution of 1/60 photon recoil. The
presented results indicate important consequences for the
production and the properties of atom laser beams. In order to
extract atom laser beams with a narrow transverse momentum spread,
the interaction of the atoms with the mean-field of the remaining
condensate has to be minimized. This might be achieved by
performing the output coupling at the very bottom of the
Bose-Einstein condensate or by reducing the condensate density. An
alternative approach may also be the extraction of atom laser
beams from an interaction-free Bose-Einstein condensate in the
vicinity of a Feshbach resonance \cite{Inouye1998}. Since the
transverse mode of the atom laser beam results from the
interaction of the atoms with a time-independent conservative
potential, one may -- in principle -- be able to compensate the
momentum spread of the beam by a suitably shaped atom optical
element. Since high quality, short focal length lenses for neutral
atoms are very difficult to realize, focussing atom lasers with a
curved mirror seems to be a promising alternative.

This work has been supported in part by DFG and the Danish
research council, BEC 2000+, and benefitted from discussions
during the KITP quantum gases program, Santa Barbara 2004.

{\it Note added in proof:}  Recently, a similar result has been
reported \cite{Riou2005}.

\end{document}